\documentclass[a4paper,11pt]{article}

\usepackage{jheppub} 

\usepackage[T1]{fontenc} 
\usepackage{graphicx}
\usepackage{epsfig}
\usepackage{color}
\usepackage{longtable}
\usepackage{hyperref}
\usepackage{latexsym}
\usepackage{amsmath}
\usepackage{amssymb}
\usepackage{dsfont}
\usepackage{verbatim}
\usepackage{bm}

\newcommand{\FC}{\;,}

\newcommand{\qbar}{\overline q}
\newcommand{\I}{\mathrm{i}}  
\newcommand{\E}{\mathrm{e}}  

\title{\boldmath Axial resonances $a_1(1260)$, $b_1(1235)$ and their decays from the lattice}

\author[a]{C. B. Lang,}
\author[b]{Luka Leskovec,}
\author[c]{Daniel Mohler,}
\author[b,d]{and Sasa Prelovsek}

\affiliation[a]{Institute of Physics,  University of Graz, A--8010 Graz, Austria}
\affiliation[b]{Jozef Stefan Institute, 1000 Ljubljana, Slovenia}
\affiliation[c]{Fermi National Accelerator Laboratory, Batavia, Illinois 60510-5011, USA}
\affiliation[d]{Department of Physics, University of Ljubljana, 1000 Ljubljana, Slovenia}

\emailAdd{christian.lang@uni-graz.at}
\emailAdd{luka.leskovec@ijs.si}
\emailAdd{dmohler@fnal.gov}
\emailAdd{sasa.prelovsek@ijs.si}

\abstract{
The light axial-vector resonances $a_1(1260)$ and $b_1(1235)$ are explored in $N_f\!=\!2$ lattice QCD by simulating the corresponding scattering channels $\rho \pi$ and $\omega\pi$.  Interpolating fields $\qbar q$ and  $\rho\pi$ or $\omega\pi$ are used to extract the $s$-wave phase shifts for the first time. The $\rho$ and $\omega$ are treated as stable and we argue that this is justified in the considered energy range and for our parameters $m_\pi\simeq 266~$MeV and $L\simeq 2~$fm. We neglect other channels that would be open when using physical masses in continuum. Assuming a resonance interpretation  a Breit-Wigner fit  to the phase shift gives the $a_1(1260)$ resonance mass  $m_{a_1}^{\textrm{res}}=1.435(53)(^{+0}_{-109})~$GeV compared to $m_{a_1}^{\textrm{exp}}=1.230(40)~$GeV. The $a_1$ width $\Gamma_{a_1}(s)\equiv g^2 p/s$ is parametrized in terms of the coupling  and we obtain $g_{a_1\rho\pi}=1.71(39)~$GeV compared to  $g_{a_1\rho\pi}^{\textrm{exp}}=1.35(30)~$GeV derived from $\Gamma_{a_1}^{\textrm{exp}}=425(175)~$MeV. In the $b_1$ channel, we find energy levels related to $\pi(0)\omega(0)$ and $b_1(1235)$, and the lowest level is found at $E_1 \gtrsim m_\omega+m_\pi$ but is within uncertainty also compatible with an attractive interaction. Assuming the coupling $g_{b_1\omega\pi}$ extracted from the experimental width we estimate $m_{b_1}^{res}=1.414(36)\genfrac{(}{)}{0pt}{}{+0}{-83}$.
}


\begin{document}
\maketitle
\flushbottom

\section{Introduction}\label{sec:introduction}

Ab-initio lattice simulations of light mesons are complicated by the fact that the majority of these states are hadronic resonances and therefore decay under the strong interaction. This is the case also for all vector ($J^{P}=1^{-}$) and axial-vector ($J^{P}=1^{+}$) mesons. The vector resonance $\rho$ is the only hadronic resonance that has been treated  on the lattice by several collaborations \cite{Aoki:2007rd,Feng:2010es,Lang:2011mn,Aoki:2011yj,Pelissier:2012pi,Dudek:2012xn}, taking into account its unstable nature and extracting its resonance mass as well as the width. First steps in this direction were also made recently for the vector resonance $K^*(892)$ \cite{Prelovsek:2013ela}, where mass and width were extracted. 

In the present paper we aim at simulating the ground state axial-vector resonances $a_1(1260)$ and $b_1(1235)$ with $I^G(J^{PC})=1^-(1^{++})$ and $1^+(1^{+-})$, respectively, taking into account their strong decays $a_1\to \rho\pi$  and $b_1\to \omega \pi$. The experimental mass and  width $\Gamma_{b_1}=142\pm 9~$MeV of the  $b_1(1235)$ are known rather precisely. The determination of the width for the broader $a_1$ from the lattice is important in view of its large uncertainty $\Gamma_{a_1}=250-600~$MeV  assigned by the  Particle Data Group  \cite{pdg12}. While most experiments and phenomenological extractions agree on the mass of the $a_1$ leading to  a PDG value of $m_{a_1}^{\textrm{PDG}}=1.230(40)~$GeV \cite{pdg12}, determinations of $\Gamma_{a_1}$ from diffractive processes where the extraction of the resonance parameters has considerable model dependence \cite{Basdevant:1977ya}, deviate substantially  from an analysis of data obtained from $\tau \rightarrow a_1 \nu$ \cite{Tornqvist:1987ch}. As an example for data obtained from a diffractive process, a recent COMPASS measurement published in 2010 \cite{PhysRevLett.104.241803} provides a much smaller error-bar $\Gamma_{a_1}=367\pm 9{\genfrac{}{}{0pt}{}{+28}{-25}}~$MeV. 

Our simulation is the first attempt at extracting the $s$-wave scattering  phase shifts for $\rho\pi$ or $\omega\pi$ channels in lattice QCD. We are interested in the energy region covering the lowest resonance states, i.e., up to $\mathcal{O}(1.5)$ GeV. To determine the phase shift in the $a_1$ channel, for example, we first extract the  discrete energy levels of the system with $I^G(J^{PC})=1^-(1^{++})$ and total momentum zero.  These levels are related to the resonant state $a_1$  as well as the two-particle states $\rho(\mathbf{ p})\pi(-\mathbf{ p})$  with $\mathbf{p} \simeq \tfrac{2\pi}{L}\mathbf{ n}$. Both, $\qbar q$ as well as meson-meson interpolators are implemented. The energy levels render the scattering phase shifts via  L\"uscher's relation  \cite{Luscher:1985dn,Luscher:1986pf,Luscher:1990ux,Luscher:1991cf} and the energy-dependence of the phase shift allows the determination of the resonance mass as well as the width.

Within this simulation we assume the $\rho$ and $\omega$ to be stable. Furthermore, additional decay channels of the $a_1$ and $b_1$ are neglected. Our assumptions and their justification are addressed in section \ref{sec:assumptions}.

The decay $b_1\to \omega \pi$ was simulated on the lattice  earlier only by McNeile and Michael   \cite{McNeile:2006bz}, where the Wick contractions  with back-tracking loops (i.e., quark lines running from source to source or from sink to sink) were omitted. The width $\Gamma[b_1\to \omega \pi]$ was extracted  based on the amplitude method \cite{McNeile:2002az} and reasonable agreement with experiment was found (see figure 3 of \cite{McNeile:2006bz}).  The $a_1$ has never been simulated taking into account its strong decay, with exception of our preliminary results \cite{Prelovsek:2011im}. 

While in quantum field theory with dynamical quarks one expects to find contributions of the intermediate meson-meson states even in correlators of $\qbar q$, in practice such signals were not clearly observed. The masses of $a_1$ and $b_1$ were previously determined by a number of lattice collaborations within the so-called single-hadron approach. In this approach only $\bar qq$ interpolators are used. One assumes that the coupling to two-particle interpolators of the type $V\pi$ ($V=\rho,\omega$) is negligible and that the mass equals the observed ground state energy level  $m\!=\!E$ (for $P\!=\!0$).  It turned out to be a better strategy to include meson-meson interpolators explicitly in the set of source and sink operators and to study the meson-meson system this way. Indeed in this more complete, coupled system level shifts as compared to the non-interaction case can be observed. The assumption that two-particle interpolators may be neglected is particularly questionable in our 
 situation, since $V(0)\pi(0)$ with $E\simeq m_V+m_\pi < m_{a_1,b_1}$ is the ground state of the system.

The $a_1$ and $b_1$ mesons have been extensively studied also by non-lattice methods.  Weinberg derived $m_{a_1}/m_{\rho}\simeq \sqrt{2}$  based on spectral functions of currents \cite{Weinberg:1967kj}. More recently, axial mesons  emerged as dynamically generated resonances in the study of scattering between pseudoscalar and vector mesons based on chiral Lagrangians \cite{Lutz:2003fm,Wagner:2007wy,Wagner:2008gz}. Unitarized Chiral Perturbation Theory models render poles related to axial mesons dynamically although these are not present in the original Lagrangian \cite{Roca:2005nm}.  This approach was used also to determine the dependence of their masses and widths as a function of number of colors ($N_c$). It was found that this dependence significantly deviates from the expectation $M\propto N_c^0$ and $\Gamma\propto 1/N_c$ for pure $\qbar q$ structure \cite{Geng:2008ag}. The chiral properties and issues related to chiral restoration were considered for example in \cite{Glozman:2007ek,Glozman:2012fj}.  

After introducing our lattice setup, we address the assumptions made within this study. Afterwards we present the masses of the scattering particles $\pi$, $\rho$ and $\omega$ in section \ref{sec:single_particles}. The interpolators and the energy spectra in the $a_1$ and $b_1$ channels are discussed in section \ref{sec:E}, while the  Wick contractions are relegated to the appendix. The analysis related to  the scattering phase shifts and the resonance parameters is given in two subsequent sections followed by our conclusions.

 \section{Lattice setup}\label{sec:lattice}
 
The  simulation is based on one ensemble of $N_f\!=\!2$ gauge configurations with clover Wilson $u/d$ quarks, generated by the authors of \cite{Hasenfratz:2008ce,Hasenfratz:2008fg}. The dynamical and valence $u/d$ quarks have the same mass corresponding to $m_\pi\!=\! 266(4)~$MeV. The parameters of the ensemble are shown in  table \ref{tab:lattice}, while more details are given in \cite{Lang:2011mn}. Due to the limited data for just a single ensemble, our determination of the lattice spacing $a$ reported in \cite{Lang:2011mn} results from taking a typical value of the Sommer parameter $r_0$. The uncertainty associated with this choice might lead to a small shift of all dimensionful quantities.   

The sea and valence quarks obey periodic boundary conditions in space. Valence quark propagators periodic and anti-periodic in time are combined into so-called ``$P+A$'' propagators, which effectively extends the time direction to $2N_T=64$ \cite{Lang:2011mn}. We use translation invariance in $t$ to sum over correlators from sources in every time slice. Errors on all values provided are obtained from single-elimination jack-knife procedure. For each jack-knife subset the full analysis of eigenstates and the subsequent evaluations are done giving the statistical variance of the results.

The rather small volume $V=16^3\times 32$ ($L\simeq 2~$fm) simplifies the use of the powerful full distillation method \cite{Peardon:2009gh}, which allows for the computation of all contractions for the correlation matrix with $\qbar q$ and $V\pi$  interpolators. 
A small box also has the advantage that the effect of $\rho\to 2\pi$ and $\omega \to 3\pi$ is less significant in our simulation with total momentum zero. 

\begin{table}[tb]
\centering
\begin{tabular}{|cccccc|}
\hline
$N_L^3\times N_T$ & $\beta$ & $a$[fm] & $L$[fm] & \#cfgs & $m_\pi$[MeV] \\ 
\hline
$16^3\times32$ & 7.1 & 0.1239(13) & 1.98 & 280 & 266(4) \\
\hline
\end{tabular}
\caption{\label{tab:lattice}Parameters of the $N_f\!=\!2$ gauge configurations  \cite{Lang:2011mn}.}
\end{table}

\section{Discussion of assumptions}\label{sec:assumptions}

In this section we discuss the assumptions in our study of two channels at hand. We assume that for $E< 1.5$ GeV the $a_1$ and $b_1$ channels are dominated  by $\rho\pi$ and $\omega\pi$ scattering in $s$-wave, and that $\rho$ and $\omega$ are stable. Therefore we assume that the effects of other channels, higher partial waves and $\rho/\omega$ decays do not significantly affect our conclusions.

Let us first focus on the effect of $\rho\to 2\pi$ decay in $\rho\pi$ scattering.   
In the absence of a rigorous framework for the analysis of $a_1\to 3\pi$ we omit $3\pi$ lattice interpolators. We expect that they have little effect on the energy levels in the relevant energy region. In order to address this, we first have to emphasize the role of our lattice environment, in particular the unphysical high pion mass and the effect of the lattice volume. The analysis relies on the discrete energy eigenstates of the correlators between a set of lattice operators. The relevance of an interpolator $O_i$ for an eigenstate $n$ can be estimated from its overlap $\langle O_i|n\rangle$. Allowing for additional $\pi\pi\pi$ interpolators in $1^{++}$ channel in principle allows for $\rho\to \pi\pi$ transition in $\rho\pi$ scattering. This might lead to shifts of  $1^{++}$ energy levels (compared to the uncoupled $\rho-\pi\pi$ situation) if $\pi\pi$ and $\rho$ levels are close in energy, i.e., within  $\simeq\Gamma_\rho$. We know from earlier work \cite{Lang:2011mn} that for our lattice parameters the ground state at rest $\rho(0)$  has negligible overlap with the $\pi(1)\pi(-1)$ operator \footnote{The meson momenta given in parenthesis $M(\mathbf{p})$ are in physical units or in units of $2\pi/L$.}. This is because this lowest two-pion interpolator allowed by the conservation of momentum and  angular momentum has energy $2(m_\pi^2+(2\pi/L)^2)^{1/2}\simeq 1.4\; \mathrm{GeV}\gg m_\rho$ \cite{Lang:2011mn}. The $\rho(1)$ with momentum $p=2\pi/L$ would have been more significantly affected by the  $\pi(1)\pi(0)$ decay channel due to its vicinity. In this study we restrict ourselves to  $E<1.5~$GeV and $\rho(p)$ has momentum $p<2\pi/L$ for both energy levels of interest in Table \ref{tab:a1}.
Although the $\rho\to \pi\pi$ channel is formally open as soon as $\sqrt{s_{\pi\pi}}>m_\rho$,
we expect that the $\rho\pi\to 3\pi$ decay channel does not significantly influence the two lowest energy levels in our energy region 
\begin{align}
\label{assumptions}
E< E_{\rho(1)\pi(-1)}\simeq(m_\rho^2+(\tfrac{2\pi}{L})^2)^{1/2}+(m_\pi^2+(\tfrac{2\pi}{L})^2)^{1/2}\simeq 1.7~\textrm{GeV}~.
\end{align}
In order to avoid $\rho(1)\to \pi(1)\pi(0)$ we restricted our simulation to the total momentum $P\equiv|\mathbf P|\!=\!0$, where the possible effect of vector meson decay is least significant.

Let us point out that analytical frameworks for rigorous multi-channel treatment on the lattice have been proposed, 
but their realization in practice remains a serious challenge. Analytic approaches for the scattering of unstable particles (with some emphasis on $s$-wave scattering) in the finite volume were recently formulated in  \cite{Roca:2012rx}, while the related challenging problem of three particles in a finite box was analytically considered in \cite{Hansen:2013dla,Polejaeva:2012ut,Briceno:2012rv,Briceno:2013rwa}. In Ref. \cite{Roca:2012rx} the approach was applied to study the effect of the $\rho \to \pi\pi$ decay to the  $\rho\pi$ scattering. Its effect  on the discrete spectrum at zero momentum is found to be negligible on our lattice with $L\simeq 2~$fm and $m_\pi\simeq 266~$MeV in the region of interest $E< E_{\rho(1)\pi(-1)}\simeq 1.7~\textrm{GeV}$
where $a_1$ resides \cite{Roca:2012rx}.  The effect of $\omega\to 3\pi$ decay for $\omega\pi$ scattering has not been studied in detail, but its influence is suppressed due to the narrow width $\Gamma^{exp}_\omega \simeq 8~$MeV and since at least two pions must have non-vanishing momentum when $\omega$ decays at rest.

More generally,  lattice interpolators with intrinsically higher energy have very little overlap with lower energy eigenstates and therefore little impact on these eigen-energies. So the threshold (the sum of masses) to some channels may be formally open but the actual eigenstate energy may be considerably higher. In that spirit we argue that interpolators which represent such higher energy content may be neglected when studying only low energy eigenstates. 

In addition to $\rho\pi$ and $\omega\pi$ there are further two-meson decay channels to consider. The $a_1$ can also couple to $f_0(500)\pi$ and $f_0(980)\pi$ in $p$-wave, which goes to three pions. The lowest possible interpolator with three pions combining to total spin 1 needs pions with non-vanishing momentum units like $\pi(0)\pi(1)\pi(-1)$ and thus --  for our parameters --  has an energy  clearly above the range considered here.

Another possible coupled channel for $a_1$ and $b_1$ is $K \bar K^*$ in $s$-wave; its threshold would be close to $1.4~$GeV, but it cannot be produced from $\bar ud$, $\pi\rho$ or $\pi\omega$ interpolators in our simulation with only
$N_f\!=\!2$ dynamical quarks. The $K\bar K^*$ could appear only by explicitly incorporating the corresponding interpolator
with $s$ quarks as valence quarks, but  $K\bar K^*\to \bar ud$ transition would be Zweig suppressed and we omit such interpolators.

The $b_1$ channel also couples to $\eta_2\rho$, its contribution might be relevant only above threshold for energies above $\mathcal{O}(1.5)$ GeV and we omit this interpolator.\footnote{The $\eta_2$ corresponds to $\eta$ for $N_f\!=\!2$ dynamical quarks and has a mass between the $\eta$ and the $\eta'$} To summarize, we assume that the elastic scattering $\rho\pi$ and $\omega\pi$ dominates both  channels in the energy region of interest and rely on the same assumption when extracting $\Gamma_{a_1\to \rho \pi}$ and $\Gamma_{b_1\to \omega \pi}$ from the experimental data. We note that the inelastic scattering with coupled channels has not been treated in lattice simulations yet, while  the analytic frameworks for this challenging problem were developed, for example, in \cite{Doring:2011ip,Hansen:2012tf,Briceno:2013rwa,Garzon:2013uwa}.

 \section{Masses of $\pi$, $\rho$, $\omega$}\label{sec:single_particles}

The masses of the scattering particles $\pi,~\rho,~\omega$ are needed for the position of the thresholds and  we collect them in table \ref{tab:single_particles}. We use  $m_\pi $ as determined  in \cite{Lang:2011mn}. 

The $\rho$ mass in table \ref{tab:single_particles} was extracted as  $m_\rho\!=\!E_\rho(p\!=\!0)$ in \cite{Lang:2011mn} and is indeed found close to $m_\rho^{res}$ in all simulations \cite{Aoki:2007rd,Feng:2010es,Lang:2011mn,Aoki:2011yj,Pelissier:2012pi,Dudek:2012xn}. 
The $\omega$ energy is calculated using  quark-antiquark interpolators ${\cal O}_{1-5}^{s=n}$ given by Eq. (21)  of  \cite{Lang:2011mn} with five different Dirac/space structures, while flavor is replaced with $\bar uu+\bar dd$. This approach is consistent with our approximation of treating $\rho(p)$ and $\omega(p)$ with $p\ll 2\pi/L$ as stable; this holds well for the narrow $\omega$ and is commonly applied also for the broader $\rho$.  The distillation method enables straightforward calculation of the disconnected contributions to $\omega$ and the final correlators are averaged over all initial time-slices and three polarizations. 
The disconnected contribution is small and we find $m_\omega \simeq m_\rho$ as expected. The value of $m_\omega$ follows from a 2-exponent fit in range $t=3-12$  using interpolators ${\cal O}_{1,2,3,5}^{n}$ and $t_0=2$. This is a conservative choice with a comparatively large uncertainty and is fully consistent with the result of other possible choices. 
  
\begin{table}[tb]
\centering
\begin{tabular}{|ccc|}
\hline 
$m_\pi a$ & $m_\rho a$  & $m_\omega  a$ \\
\hline 
0.1673(16) & 0.5107(40)         & 0.514(15) \\
\hline 
\end{tabular}
\caption{\label{tab:single_particles} The masses of scattering particles in lattice units with $a^{-1}\!\simeq \! 1.59~$GeV. } 
\end{table}

\begin{figure*}[htb]
\begin{center}
\includegraphics*[width=0.95\textwidth,clip]{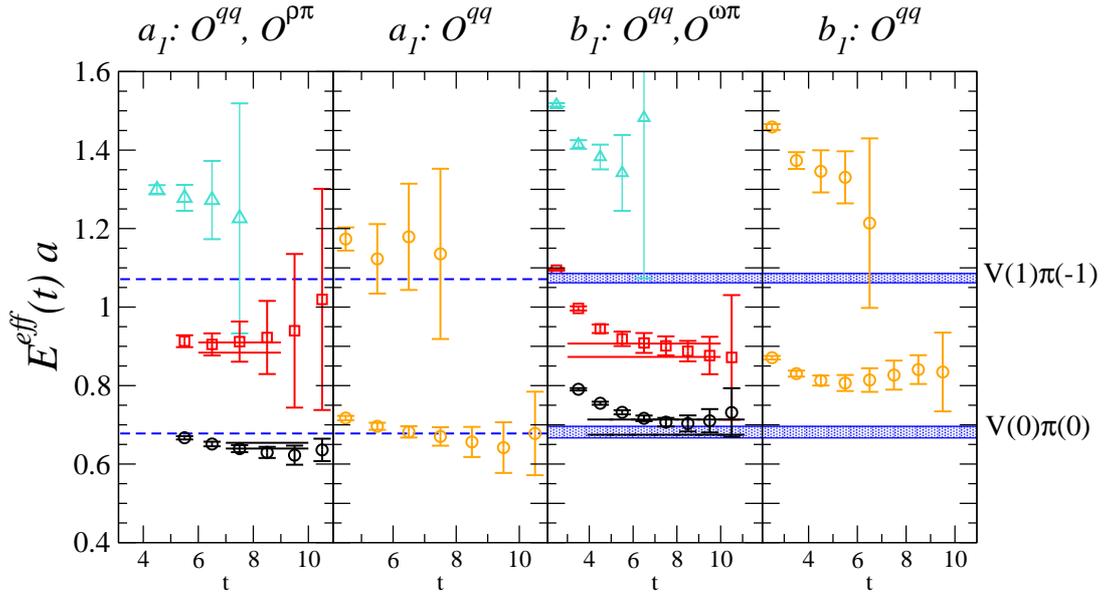}
\end{center}
\caption{Effective energies $E_n^\textrm{eff}a$ in the $a_1$ and $b_1$ channels, that correspond to the energy levels $E_na$ in the plateau region. The  horizontal lines indicate the $m_{V}+m_\pi$ threshold and the energy of a non-interacting $V(1)\pi(-1)$ state, where $V\!=\!\rho$ for $a_1$ and $V\!=\!\omega$ for $b_1$. We compare the results when ${\cal O}^{V\pi}$ is included in or excluded from the interpolator basis. }\label{fig:spectrum}
\end{figure*} 

\section{Energies in $a_1$ and $b_1$ channels}\label{sec:E}
  
 The interpolators for the $a_1$ channel with $J^{PC}=1^{++}$, 
$|a_1^-\rangle=-|I=1,I_3=-1\rangle$, $P\!=\!0$ and polarization $i$ are 
\begin{eqnarray}
\label{interpolators_a1}
{\cal O}^{\qbar q}_{1}&=&\sum _{\mathbf{x}}~\bar u(x) ~\gamma_i~\gamma_5~  d(x)\FC\\
{\cal O}^{\qbar q}_{2}&=&\sum_{\mathbf{x},j}~\bar u(x) \overleftarrow{\nabla}_j~\gamma_i~\gamma_5~ \overrightarrow{\nabla}_j ~d(x)\FC\nonumber\\
{\cal O}^{\qbar q}_{3}&=&\sum_{\mathbf{x},j,l} ~\epsilon_{ijl}  ~\bar u(x)~\gamma_j ~\tfrac{1}{2}[ \overrightarrow{\nabla}_l  -\overleftarrow{\nabla}_l ]~ d(x) \FC\nonumber\\
{\cal O}^{\rho\pi}&=&\frac{1}{\sqrt{2}}
[\pi^0(\mathbf{0})\rho^-(\mathbf{0})-\rho^0(\mathbf{0})\pi^-(\mathbf{0})]\nonumber\\
&=&\frac{1}{2}\biggl(\sum_{\mathbf{x_1}}[\bar u(x_1)\gamma_5 u(x_1) -\bar d(x_1)\gamma_5 d(x_1)]\sum_{\mathbf{x_2}}\bar u(x_2) \gamma_i d(x_2)\nonumber\\
&&\quad-\sum_{\mathbf{x_1}}[\bar u(x_1)\gamma_i u(x_1) -\bar d(x_1)\gamma_i d(x_1)]\sum_{\mathbf{x_2}} \bar u(x_2) \gamma_5 d(x_2)\biggr)\FC\nonumber 
\end{eqnarray}
where $x_i=(\mathbf{x}_i,t)$ and $\nabla$ denotes the covariant derivative. The $\rho$ and $\pi$ mesons are separately projected to zero momentum in ${\cal O}^{\rho\pi}$. 
We do not implement the interpolator $\rho(1)\pi(-1)$ (the argument $\pm 1$ indicates momenta $\pm 2\pi/L$) 
 since we concentrate on the lower energy region $E<E_{\rho(1)\pi(-1)}\simeq 1.7~$GeV.
 
Similarly, for the $b_1$ channel with $J^{PC}=1^{+-}$ and  $|b_1^-\rangle=-|I=1,I_3=-1\rangle$ we use
\begin{eqnarray}
\label{interpolators_b1}
{\cal O}^{\qbar q}_{1}&=&\sum _{\mathbf{x}}~\bar u(x) ~\gamma_i~\gamma_t~\gamma_5~  d(x)\FC\\
{\cal O}^{\qbar q}_{2}&=&\sum_{\mathbf{x},j}~\bar u(x) \overleftarrow{\nabla}_j~\gamma_i~\gamma_t~\gamma_5~ \overrightarrow{\nabla}_j ~d(x)\FC\nonumber\\
{\cal O}^{\qbar q}_{3}&=&\sum_{\mathbf{x}} ~\bar u(x)~\gamma_5 ~\tfrac{1}{2}[ \overrightarrow{\nabla}_i  -\overleftarrow{\nabla}_i ] ~d(x) \FC\nonumber\\
{\cal O}^{\qbar q}_{4}&=&\sum_{\mathbf{x}} ~\bar u(x)~\gamma_t~\gamma_5 ~\tfrac{1}{2}[ \overrightarrow{\nabla}_i  -\overleftarrow{\nabla}_i ] ~d(x) \FC\nonumber\\
{\cal O}^{\omega\pi}&=&\omega(\mathbf{0})\pi^-(\mathbf{0})= \frac{1}{\sqrt{2}}\sum_{\mathbf{x_1}}[\bar u(x_1)\gamma_i u(x_1) +\bar d(x_1)\gamma_i d(x_1)]\sum_{\mathbf{x_2}}\bar u(x_2) \gamma_5 d(x_2)\FC\nonumber
\end{eqnarray}
where, again,  $\omega$ and $\pi$ are separately projected to zero momentum. 

All quark fields $q$ in Eqs. \eqref{interpolators_a1}, \eqref{interpolators_b1} are smeared  according to the  distillation method \cite{Peardon:2009gh}, thus effectively replaced by $ \sum_{k=1}^{N_v}v^{(k)}v^{(k)\dagger}q$  with $N_v\!=\!96$, where the $v^{(k)}$ denote the Laplacian eigenvectors
of the time slice.

\begin{table*}[tb]
\setlength{\tabcolsep}{4pt}
\centering
\begin{tabular}{|cccccccccc|} 
 \hline
$n$ & $t_0$ & interp. &$\genfrac{}{}{0pt}{}{\textrm{fit}}{\textrm{range}}$ & $\tfrac{\chi^2}{\textrm{d.o.f}}$ &  $Ea$ & $\genfrac{}{}{0pt}{}{E=\sqrt{s}}{\textrm{[GeV]}}$ &  $pa$ &  $\delta~[^\circ]$ & $\tfrac{p\cot(\delta)}{\sqrt{s}}$\\
 \hline
1& 5 & ${\cal O}_{1,2}^{\qbar q},{\cal O}^{\rho\pi}$   & 7-10 & 1.1 & $0.6468(73)$& $1.030(12)$ & $\I~0.0861(95)$ &$\I~ 23(14)$ & $0.34(14)$  \\
2& 5 & ${\cal O}_{1,2}^{\qbar q},{\cal O}^{\rho\pi}$  & 6-9 & 0.015 &  $0.8977(133)$& $1.430(21)$ & $0.272(10)$ & $88.9(5.9)$ & $0.005(31)$ \\
\hline
\end{tabular}
\caption{\label{tab:a1}
Energies and phases  in the  $a_1$ channel with $I^G(J^{PC})=1^-(1^{++})$ and $P=0$, where $a^{-1}\simeq 1.59~$GeV. Both levels were obtained using a 1 exponential fit. The $p$ give the eigen-momenta of the interacting system determined from the energy levels according to Eq. \eqref{eq:E_of_p}. The ground state is below $\rho\pi$ threshold, so $p$ and $\delta$ are imaginary, while  $p\cot\delta$ is real.
}
\end{table*}

\begin{table*}[t]
\setlength{\tabcolsep}{4pt}
\centering
\begin{tabular}{|cccccccccc|}
 \hline
$n$ & $t_0$ & interp. &$\genfrac{}{}{0pt}{}{\textrm{fit}}{\textrm{range}}$ & $\tfrac{\chi^2}{\textrm{d.o.f}}$ &  $Ea$ & $\genfrac{}{}{0pt}{}{E=\sqrt{s}}{\textrm{[GeV]}}$ &  $pa$ &  $\delta~[^\circ]$ & $\tfrac{p\cot(\delta)}{\sqrt{s}}$\\
 \hline
1& 3 & ${\cal O}_{1,2,4}^{\qbar q},{\cal O}^{\omega\pi}$  & 4-11 & 0.12 & $0.694(19)$& $1.105(31)$ & $0.057(45)$ & $-3.0(6.3)$ & $-1.6(2.2)$ \\
2& 2 & ${\cal O}_{1,3}^{\qbar q},{\cal O}^{\omega\pi}$ & 3-10 & 0.049 &  $0.890(17)$& $1.418(27)$ &  $0.264(13)$ &   $93.5(7.5)$  & $-0.018(38)$   \\
\hline
\end{tabular}
\caption{\label{tab:b1}
Similar as table \ref{tab:a1} but for  $b_1$ channel with $I^G(J^{PC})=1^+(1^{+-})$. Both levels were obtained using a 2 exponential fit. The second level is consistent with threshold energy $m_\pi+m_\omega$ due to relatively large uncertainties of $E_2$ and $m_\omega$, and the corresponding phase is consistent with $\delta\simeq 0$. The  uncertainty in $m_\omega$ has negligible effect on $\delta$ for the second level.
}
\end{table*}

The energy spectrum $E_n$ is extracted from the correlation matrix 
\begin{equation}
C_{jl}(t)=\frac{1}{N_T}\sum_{t_i} \langle {\cal O}_j^\dagger (t_i+t) |{\cal O}_l(t_i)\rangle=\sum_n Z_{jn}Z_{ln}^* e^{-E_nt}
\end{equation}
averaged over all initial times $t_i$. The Wick  contractions  for both channels  are presented in figures \ref{fig:contractions_a1} and  \ref{fig:contractions_b1} of the appendix. In particular in the $b_1$ channel a rather large number of diagrams  appears.   Expressions for various elements of the correlation matrix  in terms of those Wick contractions are also provided in the appendix.  
We evaluate all the Wick contractions using the distillation method, which handles efficiently also those  with back-tracking quark lines.  

The variational method with the generalized eigenvalue equation 
\begin{equation}
C(t)\,v_n(t) =\lambda_n(t)\;C(t_0)\;v_n(t)
\end{equation} 
is applied to extract the discrete spectrum $E_n$ \cite{Michael:1985ne,Luscher:1985dn,Luscher:1990ck,Blossier:2009kd}. The resulting eigenvalues $\lambda_n(t)\sim \E^{-E_n (t-t_0)}$ give the effective energies $E_n^\textrm{eff}(t)\equiv \log [\lambda_n(t)/\lambda_n(t+1)]\to E_n$. The spectrum $E_n$ is extracted using correlated fits to $\lambda_n(t)$.

The resulting spectrum $E_n$ is shown in figure \ref{fig:spectrum}, where effective energies are plotted  for the cases when ${\cal O}^{V\pi}$ is included  or excluded from the correlation matrix. The horizontal lines indicate the position of the threshold $m_V+m_\pi$ (which has sizable uncertainty in the $b_1$ channel), and the energy of the non-interacting $V(1)\pi(-1)$ system.

We concentrate on the spectrum obtained including ${\cal O}^{V\pi}$, which is shown in the first and third pane of figure \ref{fig:spectrum} and listed in tables \ref{tab:a1} and \ref{tab:b1}. The lowest levels (circles) are near the $m_V+m_\pi$ threshold, as expected,  and its dominant component is  the $V(0)\pi(0)$ two-particle interpolator. 
The second level (squares) arises due to the presence of $a_1(1260)$ or $b_1(1235)$ resonances in these channels. The next level would be expected close to $V(1)\pi(-1)$, but we do not expect to see  it since the corresponding interpolator  is not implemented in Eqs. \eqref{interpolators_a1} and \eqref{interpolators_b1}.

The third levels in figure \ref{fig:spectrum} in both channels are noisy and unreliable, so we refrain from presenting quantitative results. Both levels correspond to masses close to $2~$GeV or above, so we see no indication for the possible existence of $a_1(1420)$, which was introduced to explain recent preliminary data by COMPASS in $a_1\to f_0\pi$ channel \cite{Paul:2013xra}. The third level in $b_1$ channel might be related to the observed $b_1(1960)$.

\section{Resonance parameters for $a_1(1260)$ }

The position of the $1^{++}$ ground state below $m_\rho+m_\pi$ threshold indicates that the energy of $\rho$ and $\pi$ is smaller if they are in the box  together than if they are in the box alone. The negative energy shift is consistent with an attractive interaction in the resonant $a_1$ channel.   
We proceed to extract the $\rho\pi$ phase shifts and the resonance parameters of $a_1(1260)$. Outside the interaction region the mesons are considered as free particles and the energy levels $E$ are related to the momenta $p$ of the two-particle state $\rho(p)\pi(-p)$  through
\begin{equation}\label{eq:E_of_p}
E=\sqrt{m_\pi^2+p^2}+\sqrt{m_\rho^2+p^2}~,
  \end{equation}
where we employ the continuum dispersion relation which applies well for the small momenta $p<2\pi/L$ of interest \cite{Lang:2011mn}.  The $s$-wave phase shifts $\delta$ for $\rho\pi$ scattering at these values of $p$ are given by the well known L\"uscher relation \cite{Luscher:1991cf}
\begin{equation}
\label{luscher}
\tan \delta(p)=\frac{\sqrt{\pi}~ p~L}{2~\mathcal{Z}_{00}\left(1;(\tfrac{pL}{2\pi})^2\right)}~,
\end{equation}
which applies above and below threshold for elastic scattering.   
The $d$-wave to  $s$-wave amplitude ratio for $a_1\to \rho \pi$ is $-0.062 \pm 0.02$ experimentally \cite{pdg12}, therefore we safely neglect the $d$-wave in Eq. \eqref{luscher}. The relation  \eqref{luscher} also neglects the contribution of the $f_0\pi$ channel, which is known to be subdominant experimentally; this simplifies one equation with several unknowns to the equation \eqref{luscher} with one unknown $\delta$ for $\rho\pi$ scattering. The $K \bar K^*$ intermediate state can not appear in our $N_f=2$ simulation with chosen interpolators, therefore the corresponding scattering parameters do not feature in the L\"uscher's relation.

The ground state below the $m_\rho+m_\pi$ threshold renders imaginary $p$ and real $p \cot\delta$ in table \ref{tab:a1}.  
The first excited level gives  $\delta \approx 90^\circ$, thus it is located close to the $a_1(1260)$ resonance mass and  $m_{a_1}\approx E_2$ holds.  

\begin{figure}[tb]
\begin{center}
\includegraphics[width=0.45\textwidth,clip]{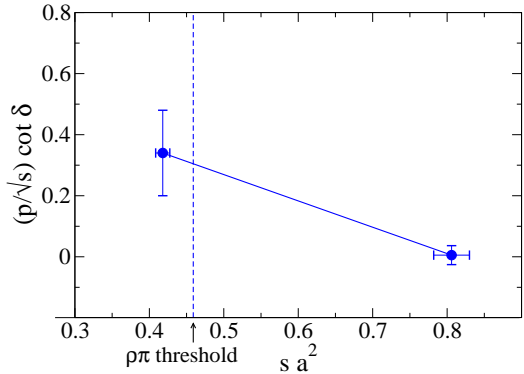}
\end{center}
\caption{We show $\frac{p}{\sqrt{s}}\,\cot \delta(s)$ with linear interpolation according to Eq.~\eqref{bw_xy}. At threshold the value is 
$[(m_\rho+m_\pi)a_{l=0}^{\rho\pi}]^{-1}$, while the position of the
zero gives the resonance mass.  
}\label{fig:bw}
\end{figure}

\begin{table*}[t]
\begin{center}
\centering
\begin{tabular}{|c | c c  c  | c c|}
 \hline
resonance     & $a_1(1260)$  & & & $b_1(1235)$ &  \\
\hline
quantity     &  $m_{a_1}^{\textrm{res}}~$ & $g_{a_1\rho\pi}~$ & $a_{l=0}^{\rho\pi}~$  &  $m_{b_1}^{\textrm{res}}~$ & $g_{b_1\omega\pi}~$ \\
     & [GeV] & [GeV] & [fm] &  [GeV] & [GeV] \\
\hline
 lat   &   $1.435(53)(^{+0}_{-109})$  & $1.71(39)$ & $0.62(28)$ & $1.414(36)(^{+0}_{-83})$  & input \\  
\hline
exp & $1.230(40)\phantom{(^{+0}_{-109})}$  & $1.35(30)$  & - &  $1.2295(32)\;\;\;\;\;\;$ & $0.787(25)$ \\  
 \hline
 \end{tabular}
\caption{The resulting Breit-Wigner masses $m^{\textrm{res}}$ together with the couplings $g$ for $a_1\to \rho \pi$ and  $b_1\to \omega \pi$, which are related to the  Breit-Wigner  width $\Gamma\equiv g^2p/s$. For the resonance masses the second uncertainty given stems from the systematic uncertainty in extracting the first excited state reliably. This uncertainty has negligible effect on the extracted coupling. The experimental values for the couplings $g$ are derived from the measured total widths \cite{pdg12} since the branching ratios to $V\pi$ have not been measured, but are expected to be largely dominant.  
The lattice value for the resonance mass of $b_1(1235)$ is obtained assuming experimental $g^{\textrm{exp}}_{b_1\omega\pi}$. All results are for our value of $m_\pi\!\simeq\! 266~$GeV. \label{tab:phy}  }
\end{center}
\end{table*}

The Breit-Wigner parametrization
\begin{equation}
\frac{-\sqrt{s}\,\Gamma(s)}{s-(m^{\textrm{res}})^2+i \sqrt{s}\,\Gamma(s)}=\frac{1}{\cot \delta-i}\FC \;\;\Gamma_{a_1}(s)\equiv g_{a_1\rho\pi}^2~\frac{p}{s}\FC  
\end{equation}
gives  $\Gamma_{a_1}(s)$  in terms of the phase space and the coupling $g_{a_1\rho\pi}$. We obtain 
\begin{equation}
\label{bw_xy}
\frac{{p}}{\sqrt{s}}\,\cot \delta(s)= \frac{1}{g_{a_1\rho\pi}^2} [ (m_{a1}^{\textrm{res}})^2 - s ]\FC
\end{equation}
which applies in the vicinity of the resonance. 
 Assuming that (like it is the case for the $\rho$) linearity \eqref{bw_xy} is a good
approximation down to the threshold and slightly below it, we 
interpolate  linearly in $s$ between the two values of $p \cot\delta/\sqrt{s}$  of table \ref{tab:a1}, as shown in figure \ref{fig:bw}. From the zero and the slope we obtain $m_{a_1}^{\textrm{res}}$ and $g_{a_1\rho\pi}$.

The resulting parameters of  the $a_1(1260)$ resonance are compared to experiment in table \ref{tab:phy}.  
The value of $m_{a1}^{\textrm{res}}$ at $m_\pi=266$ MeV is slightly higher than that of the experimental resonance $a_1(1260)$. This first lattice result for $g_{a_1\rho\pi}$ is  valuable since there is still considerable uncertainty on the total width $\Gamma_{a_1}$ and  on the $a_1\to \rho \pi$  branching ratio.  We provide the upper limit for $g^{\textrm{exp}}_{a_1\rho\pi}$  resulting from  the total width $\Gamma^{\textrm{exp}}_{a1}=250-600$ MeV \cite{pdg12}\footnote{More precisely we assume $\Gamma_{a_1}=425(175)~\mathrm{MeV}$.}, which agrees with our  $g_{a_1\rho\pi}$ within the large experimental and theoretical uncertainties.  Our lattice result for $g_{a_1\rho\pi}$ is also close to  the value $g_{a_1\rho\pi}^{phen}\approx 0.9$ GeV obtained using an Unitarized Effective Field Theory approach \cite{Roca:2005nm} (converted to our convention).

The scattering length $a_{l=0}^{\rho \pi}$ in table \ref{tab:phy} is obtained using the effective range fit $p \cot(\delta)=\tfrac{1}{a_0}+\tfrac{1}{2} r_0 p^2$ through two energy levels. Our result at $m_\pi\!=\!266~$MeV is close to $a_{l=0}^{\rho\pi}(m_\pi^{phys})\approx 0.37$ fm  obtained from Unitarized Effective Field Theory  \cite{oset_roca_private}, while  the corresponding experimental value is not known.

\section{Analysis of the  $b_1$ channel}

The robust features of the $b_1$ spectrum in figure \ref{fig:spectrum} and in table \ref{tab:b1} go along with the expectations: the  level $\omega(0)\pi(0)$ near the $m_\omega+m_\pi$ threshold, the next level close to $b_1(1235)$ with $\delta\simeq 90^\circ$ and the third level in vicinity of  $b_1(1960)$. 

However the exact position of the central value for the ground state $E_1$ in figure \ref{fig:spectrum} with respect to the threshold $m_\omega +m_\pi$ shows a slight disagreement with the expectation. It is expected to be slightly below threshold due to an attractive interaction in the resonant channel. Since $\Gamma_{b_1}<\Gamma_{a_1}$ one expects a smaller size of the energy shift $\Delta E_1=E_1-m_V-m_\pi$ in the $b_1$ channel than in the $a_1$ channel, rendering a lattice extraction challenging. We estimate the expected energy shift $a\Delta E_1 \simeq -0.01$ based on $g^{\textrm{exp}}_{b_1 \omega \pi}$, Breit-Wigner dependence \eqref{bw_xy}, the L\"uscher relation \eqref{luscher} and the value of  $m^{\textrm{res}}_{b_1}$ \eqref{mr_b1} determined below; note that this shift is smaller than the uncertainty of the ground state energy level in the $b_1$ channel and comparable to the uncertainty in $am_\omega$.  A correlated analysis reveals that our ground state $E_1$ is compatible with $m_\omega+m_\pi$ within the large uncertainties, although the central value leads to $E_1 \gtrsim m_\omega+m_\pi$. We stress that this discrepancy with expectations is not statistically significant.

If upon improving the statistics the ground state level still stays above threshold, this would be hard to understand. The three-pion state, for example,  could hardly explain such behavior since the lowest state $\pi(0)\pi(1)\pi(-1)$ with $J\!=\!1$  has energy $\simeq\! 1.6~$GeV, which  is significantly above the threshold on our lattice. 

In the $b_{1}(1235) \to \omega \pi$ channel the d-wave to s-wave amplitude ratio observed in experiment is $0.277 \pm 0.027$ \cite{pdg12}, and the d-wave contribution, which we neglect in Eq. \eqref{luscher}, might play an important role as it for example does in the deuteron.

Due to the discussed uncertainty of the ground state energy level we determine the resonance $b_1(1235)$ mass using the Breit-Wigner relation \eqref{bw_xy},
 \begin{equation}
 \label{mr_b1}
 m_{b_1}^{\textrm{res}}=\biggl[E_2^2+(g^{\textrm{exp}}_{b_1\omega \pi})^2\biggl(\frac{p \cot\delta}{\sqrt{s}}\biggr)^2\biggr]^{1/2}
=1.414(36)(^{+0}_{-83})~\mathrm{GeV}\;,
 \end{equation}
 based on $E_2$ and $(p\cot\delta/\sqrt{s})_2$ for the second level in table \ref{tab:b1}, while we assume 
 the experimental value of the coupling $g^{\textrm{exp}}_{b_1\omega\pi} =0.787(25)~$GeV. It is  derived from $\Gamma^{\textrm{exp}}_{b_1}$ assuming $Br[b_1\to \omega \pi]\simeq 1$,  which has not been measured  but is expected to be valid to a good approximation. The resulting resonance mass at our $m_\pi$ is somewhat higher than the experimental one. 
 
 \section{Conclusions and outlook}
 
We presented the first lattice simulation of light axial resonances $a_1$ and $b_1$ taking into account their dominant strong decays modes $\rho\pi$ and $\omega\pi$. The interpolating fields $\qbar q$ as well as $V\pi$ ($V=\rho,\omega$) were used for this purpose. We find an energy level near the $V(0)\pi(0)$ threshold and excited levels close to the resonance positions of $a_1(1260)$ and $b_1(1235)$. The $s$-wave phase shifts for $V\pi$ scattering were extracted using the L\"uscher relation and  they are fitted using the Breit-Wigner formula. 
The  resonance parameters are compared to experiment  in table \ref{tab:phy}.  

The resonance masses $m_{a_1}^{\textrm{res}}$ and $m_{b_1}^{\textrm{res}}$ at our $m_\pi\simeq 266~$MeV are somewhat higher than the experimental values. The $a_1\to \rho\pi$ coupling $g_{a_1 \rho \pi}$, which parametrizes the corresponding decay width, agrees with experiment within sizable error bars. The analogous $b_1\to \omega\pi$ coupling was not extracted since the central value for the ground state in the $b_1$ channel is found slightly above  $m_\omega+m_\pi$, while one expects it to be slightly below threshold in an attractive channel. 
The ground state is still consistent with $m_\omega+m_\pi$ and an attractive interaction within the sizable uncertainty. Future lattice studies with improved statistical accuracy could clarify this issue by resolving smaller energy shifts which should lead to a determination of $g_{b_1 \omega\pi}$ from Lattice QCD. 
  
Our analysis assumes stable scattering  particles $\rho$ and $\omega$, which is a reasonable approximation for our simulation with zero momentum, $L\!\simeq\! 2~$fm  and $m_\pi\!\simeq \! 266~$MeV.  However, the effects of vector meson decays might have to be taken into account in future simulations with 
physical  pion masses, multiple box sizes and possibly non-zero total momenta.
Proper treatment of this problem appears considerably more challenging. 
An analytic framework for the scattering of unstable particles in finite volume has been recently proposed in \cite{Roca:2012rx} while the related problem of three particles where two of them can resonate was considered in \cite{Hansen:2013dla,Polejaeva:2012ut,Briceno:2012rv,Briceno:2013rwa}. These approaches can serve as a guideline for future lattice simulations.

\appendix
\section{Wick contractions and the correlation matrix}\label{app}

\subsection{The $\rho\pi$ $s$-wave correlation matrix}

\begin{figure}[t!]
\begin{center}
\includegraphics[width=0.3\textwidth]{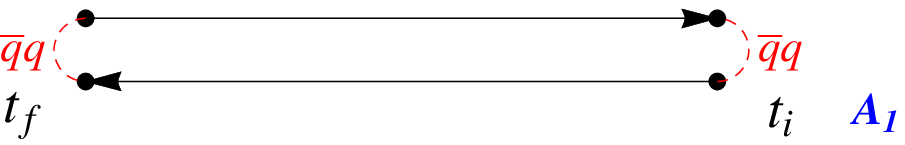}\hfill
\includegraphics[width=0.3\textwidth]{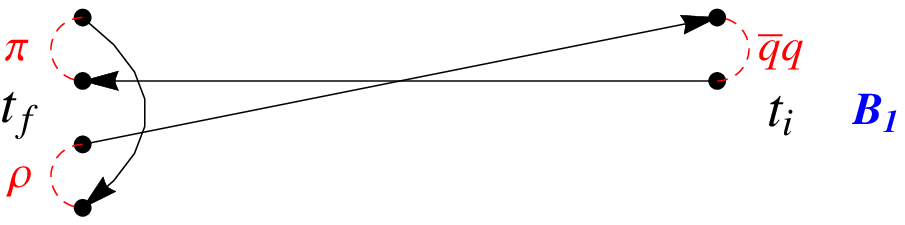}\hfill
\includegraphics[width=0.3\textwidth]{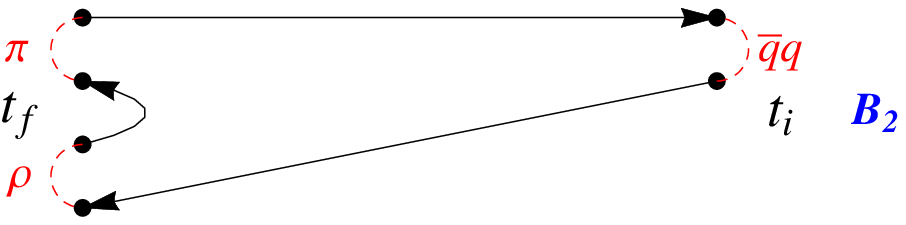}\vspace*{6pt}\\ \hfill
\phantom{\includegraphics[width=0.3\textwidth]{a1_A1.eps}}
\hfill\includegraphics[width=0.3\textwidth]{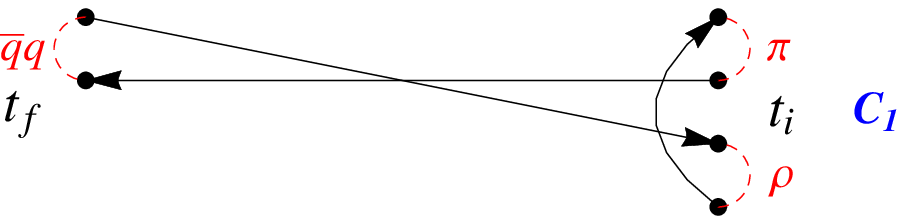} \hspace*{6pt}\hfill
\includegraphics[width=0.3\textwidth]{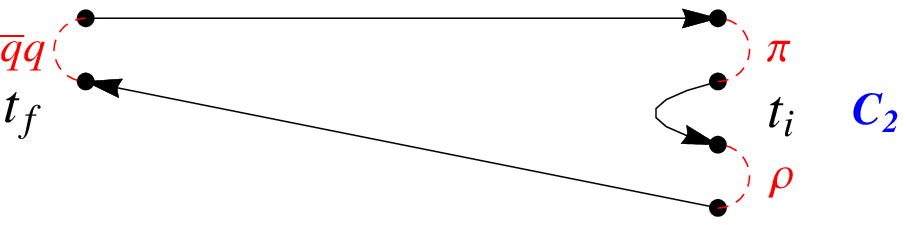}\vspace*{6pt}\hfill\\
\includegraphics[width=0.3\textwidth]{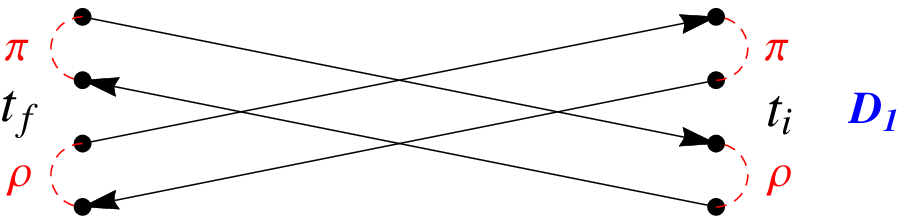}\hfill
\includegraphics[width=0.3\textwidth]{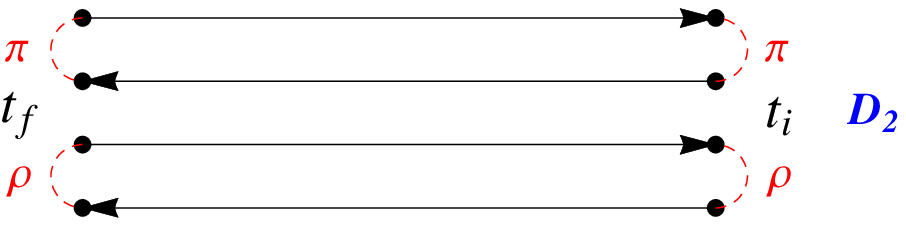}\hfill
\includegraphics[width=0.3\textwidth]{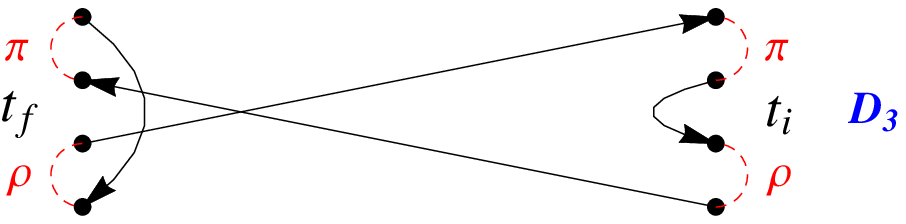}\vspace*{6pt}\\
\includegraphics[width=0.3\textwidth]{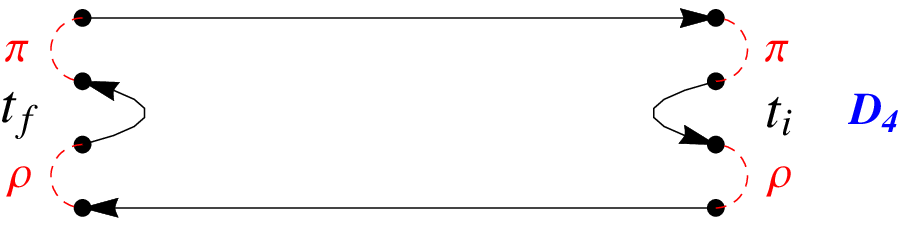}\hfill
\includegraphics[width=0.3\textwidth]{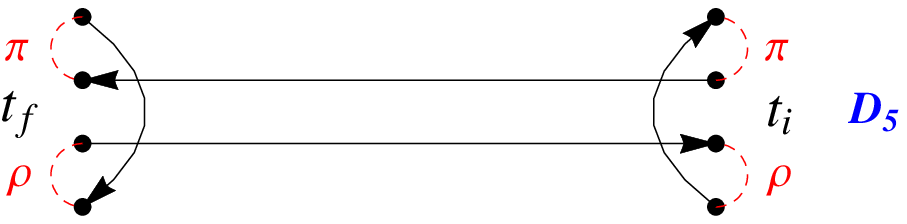}\hfill
\includegraphics[width=0.3\textwidth]{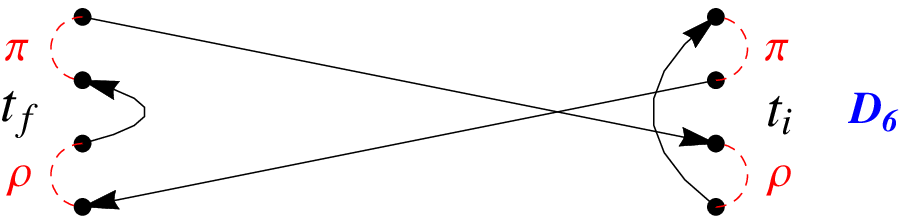}
\caption{Wick contractions in the $a_1$ channel with ${\cal O}^{\qbar q}$ and ${\cal O}^{\rho \pi}$ interpolators. 
}\label{fig:contractions_a1}
\end{center}
\end{figure}

The correlation matrix is built from four operators ${\cal O}^{\qbar q}_{1-3}$ and ${\cal O}^{\rho \pi}$ listed in Eq. \eqref{interpolators_a1}. The Wick contractions are listed in figure  \ref{fig:contractions_a1}, where solid lines denote $u/d$ quarks, while dashed lines denote the momentum projections and Dirac structures at source and sink. The correlation matrix $C(t_f,t_i)$ for the $a_1$ channel is built from the following linear combinations of the Wick contractions
\begin{align}
\label{c_a1}
\langle {O}^{\qbar q}(t_f) \mathcal{O}^{\bar{q}q \dagger} (t_i)\rangle  &= - A_{1}\FC\\
\langle {O}^{\rho \pi}(t_f) \mathcal{O}^{\bar{q}q \dagger} (t_i)\rangle  &=  B_{1} - B_{2}\nonumber\FC\\
\langle {O}^{\qbar q}(t_f) \mathcal{O}^{\rho\pi \dagger} (t_i)\rangle   &=  C_{1} - C_{2}\nonumber\FC\\
\langle {O}^{\rho \pi}(t_f) \mathcal{O}^{\rho\pi \dagger} (t_i)\rangle &= - D_{1} + D_{2} + D_{3} - D_{4} -D_{5} +  D_{6}~,\nonumber
\end{align} 
where $A_l,~B_l,~C_l,~D_l$ refer to labels in figure  \ref{fig:contractions_a1}.

\begin{figure}[t!]
\begin{center}
\includegraphics[width=0.3\textwidth]{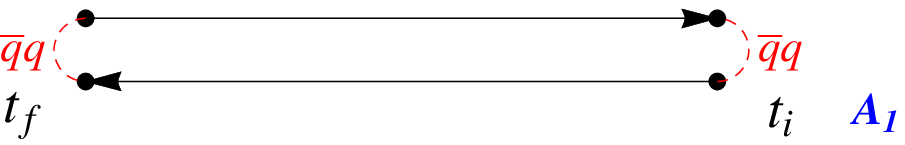}\vspace*{6pt}\\
\includegraphics[width=0.3\textwidth]{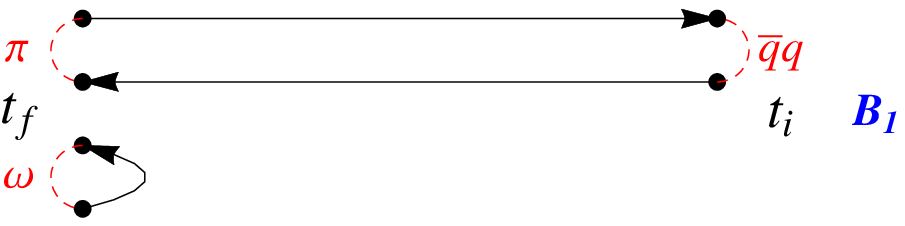}\hfill
\includegraphics[width=0.3\textwidth]{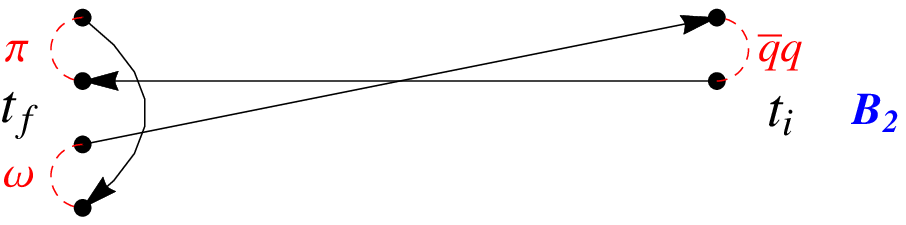}\hfill
\includegraphics[width=0.3\textwidth]{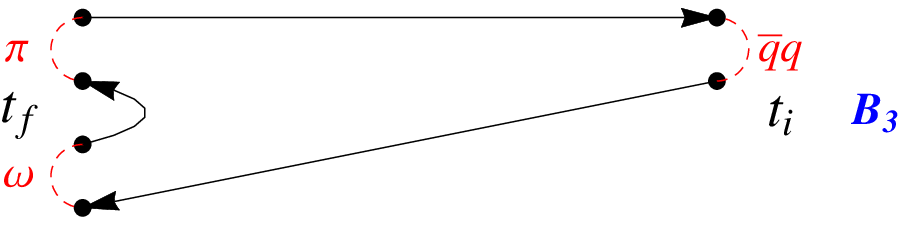}\vspace*{6pt}\\
\includegraphics[width=0.3\textwidth]{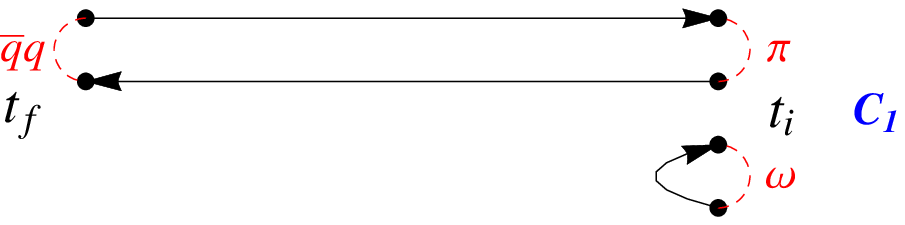}\hfill
\includegraphics[width=0.3\textwidth]{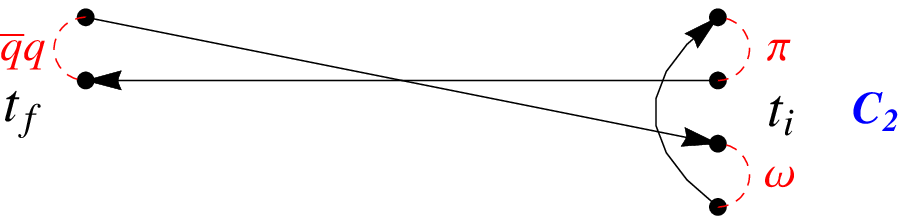}\hfill
\includegraphics[width=0.3\textwidth]{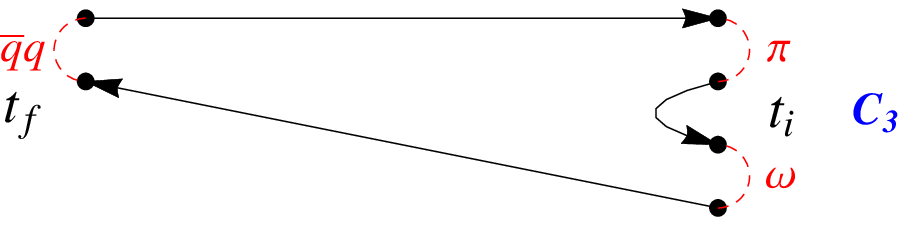}\vspace*{6pt}\\
\includegraphics[width=0.3\textwidth]{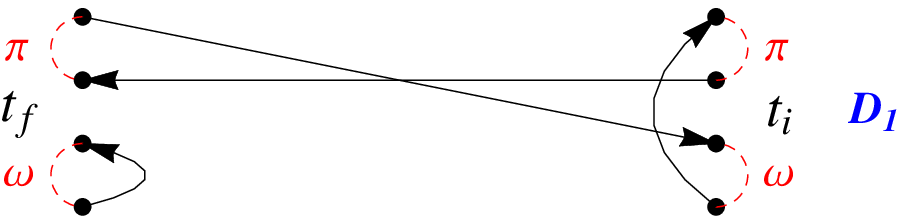}\hfill
\includegraphics[width=0.3\textwidth]{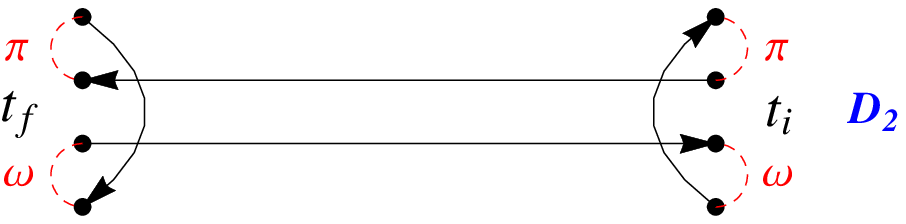}\hfill
\includegraphics[width=0.3\textwidth]{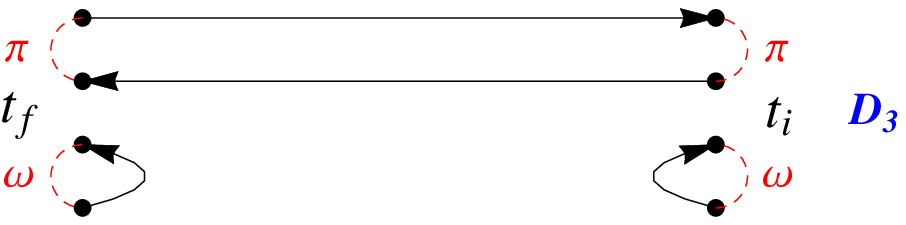}\vspace*{6pt}\\
\includegraphics[width=0.3\textwidth]{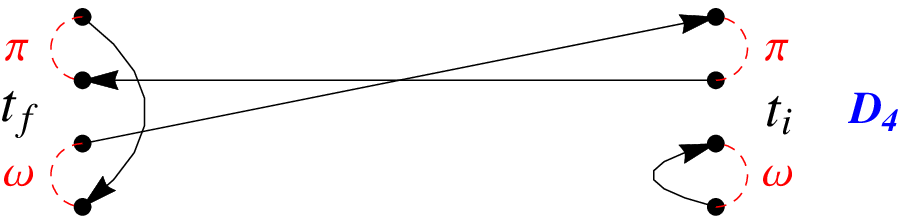}\hfill
\includegraphics[width=0.3\textwidth]{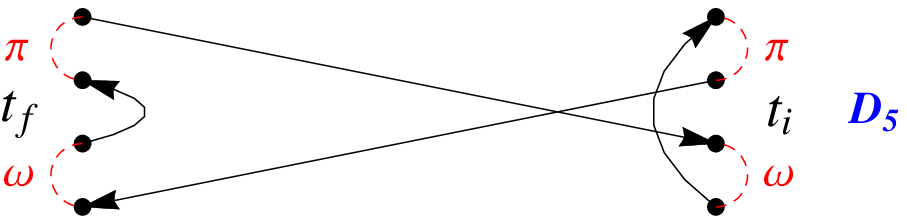}\hfill
\includegraphics[width=0.3\textwidth]{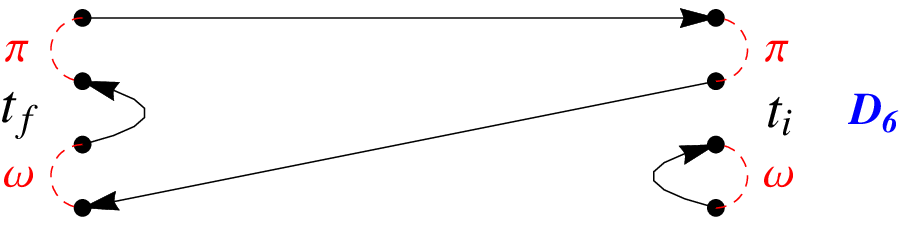}\vspace*{6pt}\\
\includegraphics[width=0.3\textwidth]{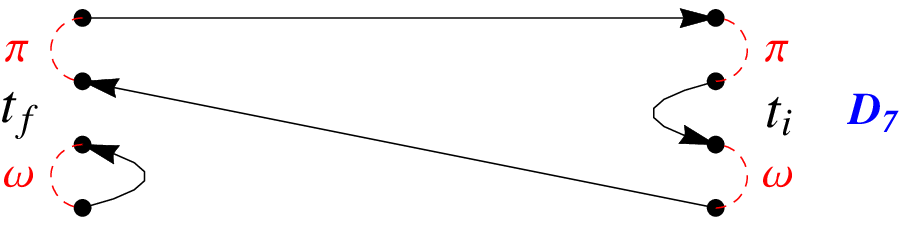}\hfill
\includegraphics[width=0.3\textwidth]{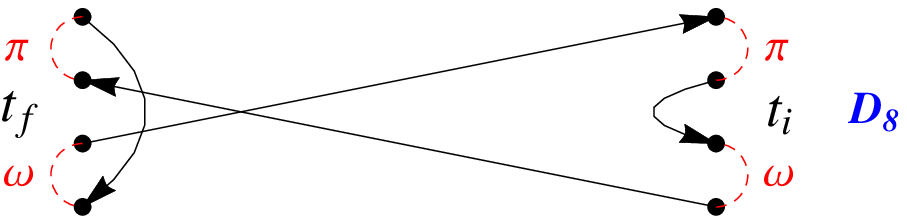}\hfill
\includegraphics[width=0.3\textwidth]{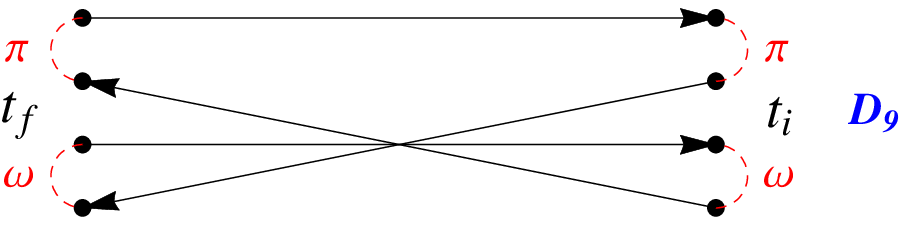}\vspace*{6pt}\\
\includegraphics[width=0.3\textwidth]{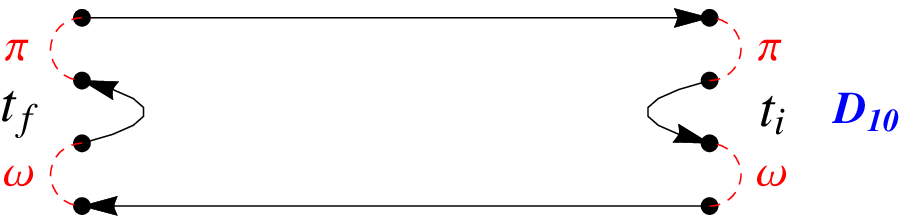}\hfill
\includegraphics[width=0.3\textwidth]{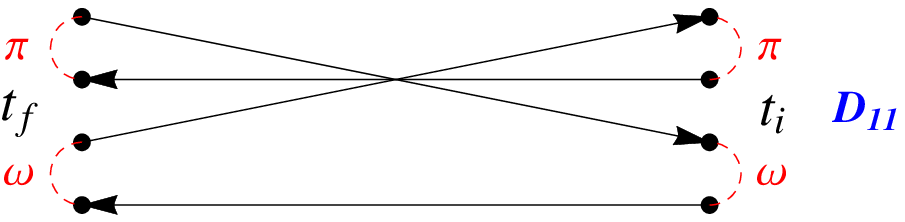}\hfill
\includegraphics[width=0.3\textwidth]{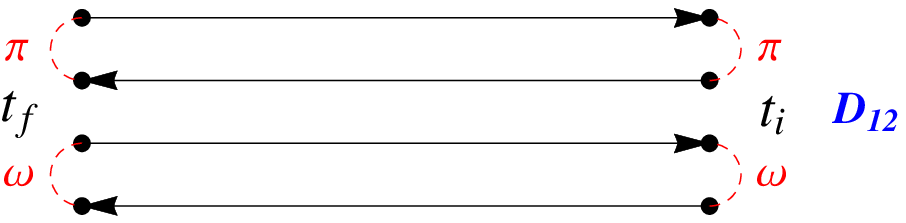}
\caption{Wick contractions in the $b_1$ channel with ${\cal O}^{\qbar q}$ and ${\cal O}^{\omega \pi}$ interpolators.}\label{fig:contractions_b1}
\end{center}
\end{figure}
\subsection{The $\omega\pi$ $s$-wave correlation matrix}

This correlation matrix is built from five operators ${\cal O}^{\qbar q}_{1-4}$ and ${\cal O}^{\omega \pi}$ listed in \eqref{interpolators_b1}. In this case there is a larger number of Wick contractions, all drawn in figure \ref{fig:contractions_b1}.  The  correlation matrix  is built from these Wick contractions as follows
\begin{align}
\label{c_b1}
\langle {O}^{\qbar q}(t_f) \mathcal{O}^{\bar{q}q \dagger} (t_i)\rangle  &= - A_{1}\FC\\
\langle {O}^{\omega \pi}(t_f) \mathcal{O}^{\bar{q}q \dagger} (t_i)\rangle  & =  \sqrt{2} B_{1} -\frac{1}{\sqrt{2}} B_{2} - \frac{1}{\sqrt{2}} B_{3}\FC\nonumber\\
\langle {O}^{\qbar q}(t_f) \mathcal{O}^{\omega\pi\dagger} (t_i)\rangle  & =  \sqrt{2} C_{1} -\frac{1}{\sqrt{2}} C_{2} - \frac{1}{\sqrt{2}} C_{3}\nonumber\FC\\
\langle {O}^{\omega \pi}(t_f) \mathcal{O}^{\omega\pi \dagger} (t_i)\rangle  & = + D_{1} -\frac{1}{2} D_{2} - 2 D_{3} + D_{4} - \frac{1}{2} D_{5} \nonumber\\
&\quad
+ D_{6} + D_{7} - \frac{1}{2} D_{8} - \frac{1}{2} D_{9} -\frac{1}{2} D_{10} - \frac{1}{2} D_{11} + D_{12}~.\nonumber
\end{align}

\acknowledgments
We thank Anna Hasenfratz for providing the gauge configurations used for this work. We are grateful for valuable discussions with  M. D\"oring, E. Oset, J. Pelaez, L. Roca and A. Rusetsky. This work is supported by the Slovenian Research Agency. Fermilab is operated by Fermi Research Alliance, LLC under Contract No. De-AC02-07CH11359 with the United States Department of Energy. We thank INT in Seattle for hospitality.



\providecommand{\href}[2]{#2}\begingroup\raggedright\endgroup

\end{document}